\documentstyle[12pt,epsf]{article} 
 \newcommand {\bi} {\bibitem}
 \newcommand {\be} {\begin{equation}}
\newcommand {\bea} {\begin{eqnarray} \nonumber }
\newcommand {\ee} {\end{equation}}
\newcommand {\eea} {\end{eqnarray}}
 
 \newcommand {\si} {\sigma}
\newcommand {\de} {\delta}

\newcommand {\la} {\lambda}

\newcommand {\lan} {\langle}
\newcommand {\ran} {\rangle}

\def \form#1 {eq. (\ref{#1}) }
\def \parziale#1#2  {{\partial {#1} \over \partial {#2}}}

\begin{document}   

\title{Short time aging in binary glasses} 
\author{  Giorgio Parisi \\ 
Dipartimento di Fisica,
Universit\`a {\sl La  Sapienza}\\ INFN Sezione di Roma I \\ Piazzale 
Aldo Moro, Roma 00187}
\maketitle

\begin{abstract}
We present some simple computer simulations that indicate that  at short 
time aging is realized in a simple model of binary glasses.  It is interesting 
to note that modest computer simulations are enough to evidenziate this effect.  
We also find indications of a dynamically growing correlation length.  
\end{abstract}

Aging has been discovered long time ago and it is has been experimentally 
studied in great details in some materials \cite{POLI}.  It has been later 
realized that it is a quite common phenomenon in physics.  It has been the 
subject of wide theoretical investigations only recently \cite{B,CUKU,FM}.  In the 
nutshell aging predicts that the response of the system to a force that has been 
applied for a time $t$ depends on $t$ (also for very large times) when $t$ is 
comparable to the waiting time $t_{w}$, i.e.  the time at which the systems has 
been carried to conditions at which the experiment has been done.

Aging has being studied analytically in generalized spin glasses: in the 
simplest form it predicts that the correlation functions among a configuration 
at time $t_{w}$ and at time $t_{w}+t$ depend only on the ration $t/t_{w}$ in the 
limit of large times \cite{CUKU}.  This form of aging is approximately found to 
be correct in spin glasses both in experiments and in numerical simulations 
\cite{QUA,QUI,QUO}, (although some small modification may be needed).  Slightly 
different forms of aging have been proposed, e.g the scaling variable could be 
$t/t_{w}^{\mu}$ with $\mu$ near but not equal to 1.  Here we stick to the 
$t/t_{w}$ scaling and we will refer to it by the name of {\em simple aging}.

The aim of this note it to start a systematic study of aging using numerical 
simulations in real glasses.  We will show that the aging regime starts at 
relatively short times and some of its properties can be investigated with a 
modest amount of computer time.  We limit ourselves to the analysis in the 
initial time region, leaving to further more systematic investigations the study of
the behaviour  at larger times and in bigger systems.

The numerical experiment we present here is rather simple: we run a numerical 
simulation where the system starts from a fully random configuration 
(i.e. at infinite temperature).  The system is then carried (at time zero) at 
temperature $T$.  We make a photograph of the system at time $t_{w}$ and we 
compare the later evolution of the system with this reference configuration.

\begin{figure}[htbp]
    \epsfxsize=400pt\epsffile[22 206 549 549]{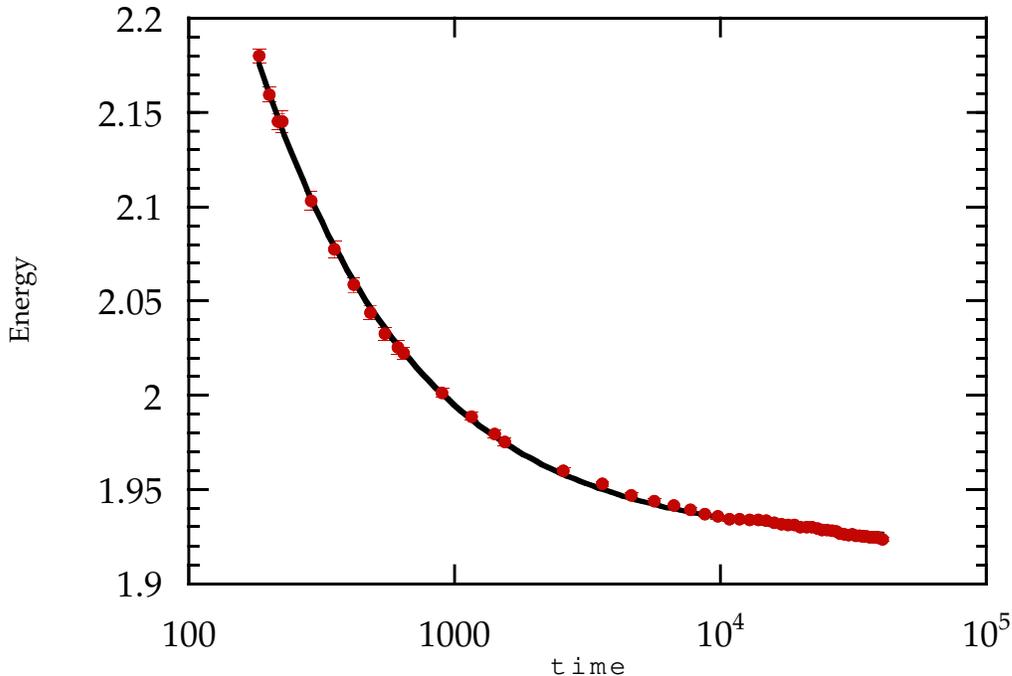} \caption{The energy 
    density for 258 particles as function of time.  The fit is 
    $E=E_\infty+At^{-\la}$ with $E_\infty=1.92$, $A=12$ and $\la=$.7.} \label{FIG1}
\end{figure}

The main quantity on which we concentrate our attention is the two times 
correlation function
\begin{equation}
g(r,t_{w},t)=\sum_{i,k=N}{ \lan \de(|x_{i}(t_{w})-x_{k}(t_{w}+t)|-r)\ran \over N},
\label{G}
\end{equation}
where $N$ is the total number of particles and $x_{i}(t)$ denotes the position 
of the particle $i$ at time $t$.  In the limit of large times $g(r,t,t)$ goes to 
the usual correlation function for liquids (apart from an extra delta function 
at the origin, which is absent in the usual definitions where the sum is 
restricted to $i\ne k$).

For this function simple  aging predicts that for {\em both} $t$ and $t_{w}$
large, we have the  scaling relation
\begin{equation}
g(r,t_{w},t)\approx G(r,s), \ \mbox{where} \ \ s\equiv \frac{t}{t_{w}}.
\label{SCALING}
\end{equation}
 
We have tried to test this relation for binary fluids.  The model we consider is 
the following.  We have taken a mixture of soft particle of different sizes.  
Half of the particles are of type $A$, half of type $B$ and the interaction 
among the particle is given by
\begin{equation}
\sum_{{i<k}} \left(\frac{(\si(i)+\si(k)}{|x_{i}-x_{k}|}\right)^{12},
\label{HAMILTONIAN}
\end{equation}
where the radius ($\si$) depends on the type of particles.  This model has been 
carefully studied studied in the past \cite{HANSEN1,HANSEN2,HANSEN3}.  It is 
known that a choice of the radius such that $\si_{B}/\si_{A}=1.2$ strongly 
inhibits crystallisation and the systems goes into a glassy phase when it is 
cooled.  Using the same conventions of the previous investigators we consider 
particles of average diameter $1$, more precisely we set
\begin{equation} 
{\si_{A}^{3}+ 2 (\si_{A}+\si_{B})^{3}+\si_{B}^{3}\over 4}=1.
\label{RAGGI}
\end{equation}
 
Due to the simple scaling behaviour of the potential, the thermodynamic 
quantities depend only on the quantity $T^{4}/ \rho$, $T$ and $\rho$ being 
respectively the temperature and the density.  For definiteness we have taken 
$\rho=1$.  The model as been widely studied especially for this choice of the 
parameters.  It is usual to introduce the quantity $\Gamma \equiv \beta^{4}$.  
The glass transition is known to happen around $\Gamma=1.45$ \cite{HANSEN2}.

Our simulation are done using a Monte Carlo algorithm, which is more easy to 
deal with than molecular dynamics, if we change the temperature in an abrupt 
way.  Each particle is shifted by a random amount at each step, and the size of 
the shift is fixed by the condition that the average acceptance rate of the 
proposal change is about .5.  Particles are placed in a cubic box with periodic 
boundary conditions and at the end of each Monte Carlo sweep all the particles 
are shifted of the same vector in order to keep the center of mass fixed
\cite{LAPA}.  This last step in introduced in order to avoid drifting of the 
center of mass and it would be not necessary in molecular dynamics if we start 
from a configuration at zero total momentum.

In our simulations we have considered a relatively small number of particles, 
$N=34$, $N=66$ and $N=258$ (most of the data we show are for $N=258$).  We start 
by placing the particles at random and we quench the system by putting it at 
$\Gamma=1.8$, i.e.  at a temperature well below the glass transition.  The 
energy as function of the Monte Carlo time ($t$) is shown in fig.  1.  The data 
are the average over 25 different realization of the dynamics with different 
initial conditions.  The energy seems to decay to an asymptotic value with some 
corrections which vanishes as a power of time.

\begin{figure}[htbp]
  \epsfxsize=400pt\epsffile[22 206 549 549]{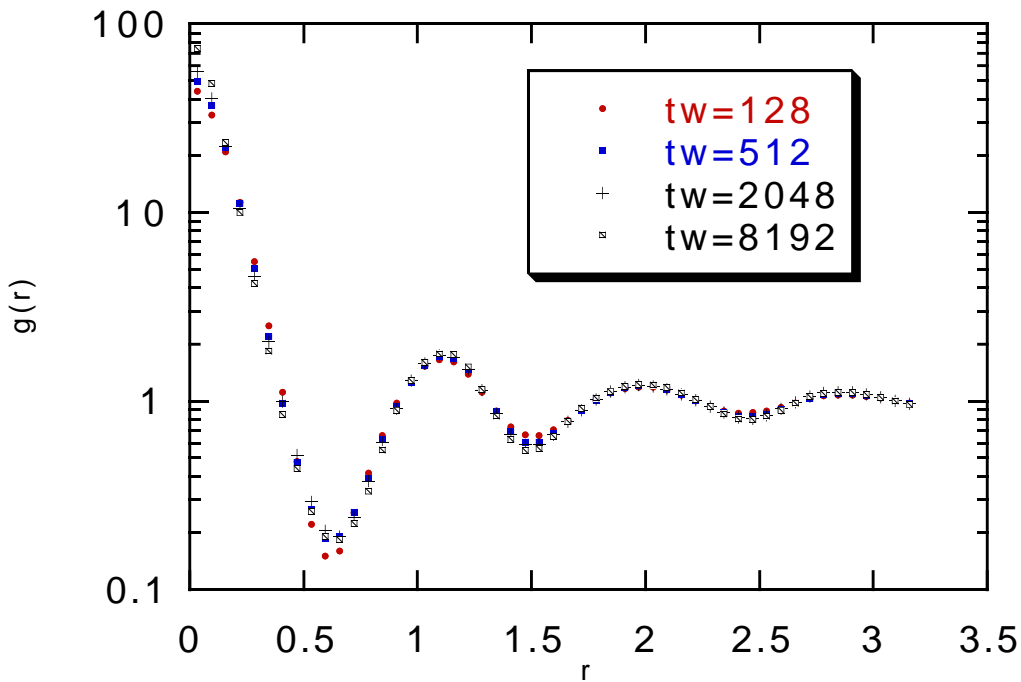} \caption{The correlation 
  function $G(r,t_w,t)$ as function of $r$ for $N=258$ for different values of 
  $t_w$ ($t_w$=128, 512,2048 ,8192, at $t/t_w$ equal to 3.  The data have been 
  averaged over 25 different initial conditions} \label{FIG2}
\end{figure}

We have measured the correlation function $G$ for different choices of $t_w$ 
(i.e. $t_w=$32, 128, 512, 2048, 8192) at $s\equiv t/t_w$ equal to 3.  If we exclude 
the points at $t_w=32$ (which we have not plotted), the data for the correlation 
functions have a very similar shape independently of $t_w$ at fixed $s$, as 
expected from simple aging.  In order to check in a more quantitative way the 
aging and to evidenziate possible violations of aging, we have introduced the 
quantity $q(t_w,t)$ defined as 
\be q(t_w,t)=\int dx g(x,t_{w},t) f(x)\equiv\sum_{i,k=1,N},
{f(x_{i}(t+t_{w})-x_{k}(t_{w})) \over N^{2}}
\ee
where we have chosen the function $f$ in such a way that it is sensitive to the 
area of the central peak, i.e.
\be f(x)={a^{12} \over x^{12}+a^{12}}, \ee 
with $a=.3$ The function $f$ is very small when $x>>.3$ and near to $1$ for 
$x<.3$.  The value of $q$ will thus be a number very near to $1$ for similar 
configurations (in which the particles have moved of less than $a$) and it will 
be much smaller  value (less than .1) for unrelated configurations; using the 
same terminology as in spin glasses \cite {EA,mpv,parisibook2} $q$ can be called 
the overlap of the two configurations.

\begin{figure}[htbp]
  \epsfxsize=400pt\epsffile[22 206 549 549]{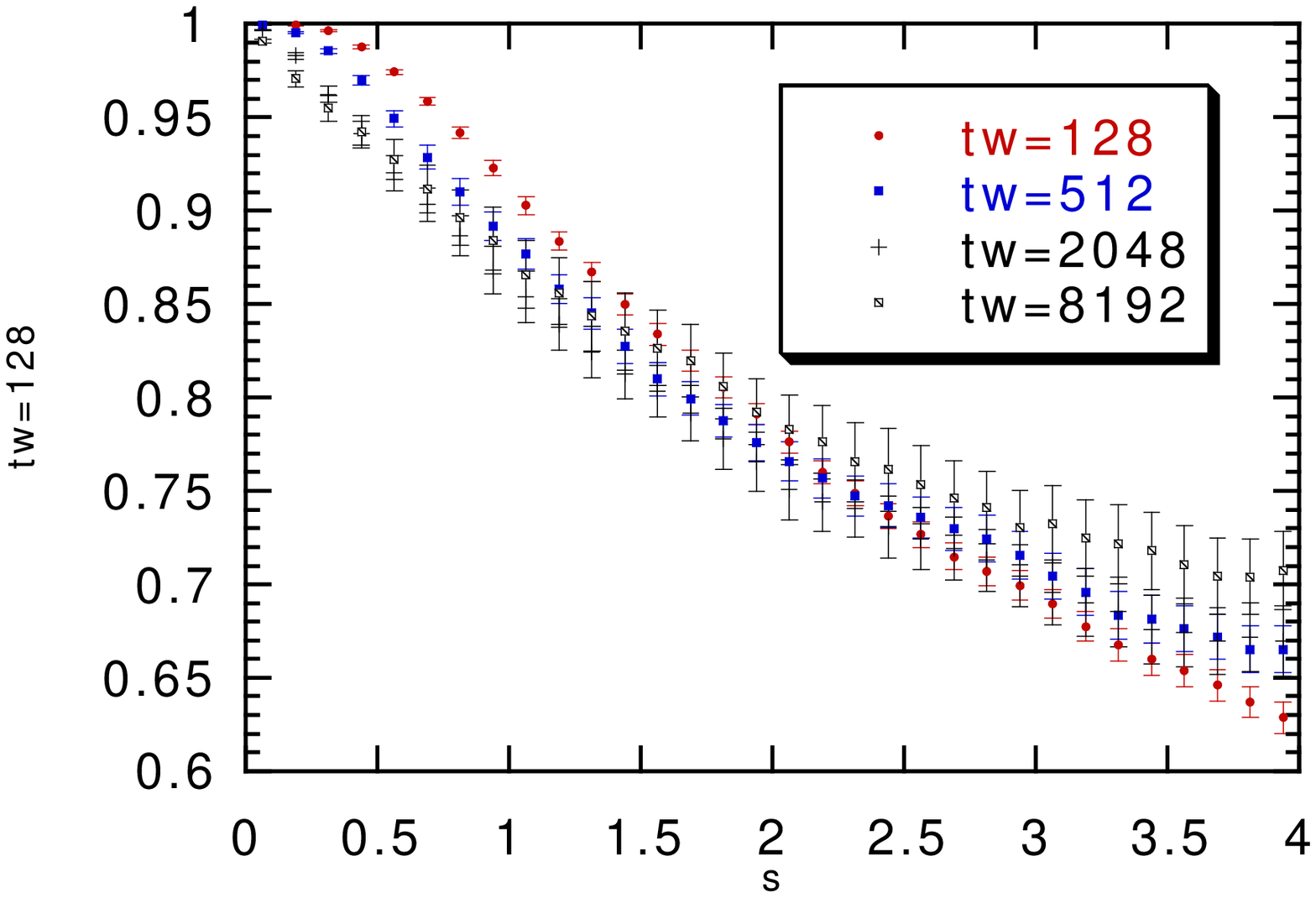}  
\caption{The overlap $q$ as function of $s$ for $t_w$=128, 512, 2048, 8192} and
$N=258$, averaged over 25 initial conditions.
 \label{FIG3} \end{figure}

In fig.3 we plot the overlap as function of $s$ at different value of the 
waiting time.  We notice that for $s$ near to zero there is noticeable 
dependence on the waiting time, which can be related to the fact that $t$ is not 
large.  There is a small drift of the data at large s.  In order to see 
evidenziate the effect we plot in fig.(4) the function $q$ at $s=4$ as function 
of $t_w$.
\begin{figure}[htbp]
  \epsfxsize=400pt\epsffile[22 206 549 549]{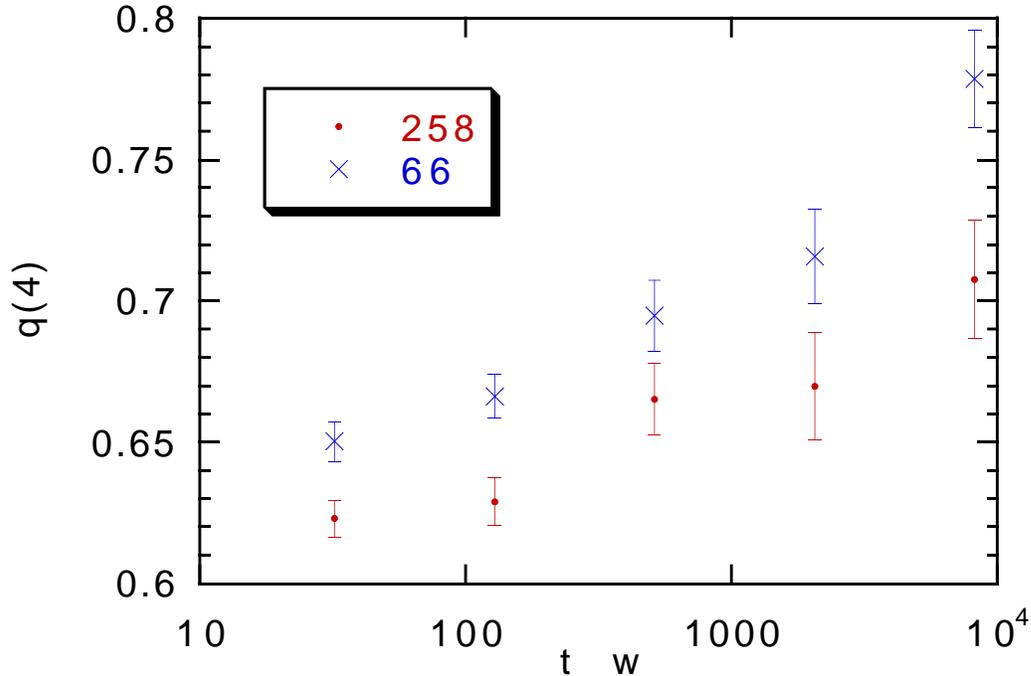}
\caption{The value of the overlap $q$ at $s=4$ as function of $t_w$} for $N=66$
and $N=258$.  
 \label{FIG4} \end{figure}

It is not clear if the increase at larger values of $t_w$ is due to finite size 
correction or to small violations of the simple aging hypothesis.  This point 
should be carefully investigated in the future and its clarification goes beyond 
the aim of this note.  The data on smaller systems indicates that there are 
definitely strong volume effects at large times \cite{BK}.  This may seem
surprising; in order to find the physical origine of this effect  it is instructive
to look to the time dependance of the fluctuations of $q$ from  sample to sample,
i.e.  to \be
\Delta (t_{w},t)\equiv<(q-<q>)^2>
\ee
where the average is done over different realization of the initial conditions.  
This quantity seems to increase a a power of $t_w$ with and exponent $\mu$ near 
to $1/2$.  In fig.  5 we show the data (at $s=4$) for 
\be
M(t_{w})\equiv N \Delta(t_{w},4t_{w})
\ee
 versus $t_w^{1/2}$.  The quantity $M(t_{w})$ seem to increase a a power of time 
 and shows a weak dependance on the size (as expected, usually fluctuations goes 
 to zero as $N^{-1/2}$).

In order to interprete this results it is convenient to recall the physical 
picture at the base of simple aging: the system evolves by a sequence of quasi 
equilibrium states and remains in a given state for a time proportional to time 
needed to arrive to it.  In the most extreme picture we have a punctuated 
equilibrium of long period of stasis intermixed by fast, thermally activated, 
tunnelling events.  Increasing the value of $t_w$ the barrier that we have to 
cross become higher ad higher and such collective movements involves a large 
number of particles.  Roughly speaking we expect that the variance $\Delta$ is 
inversely proportional to the number ($N/M(t_{w})$) of regions which have been 
moved independently, i.e.
\be
\Delta(t_w) \propto {M(t_{w}) \over N}.
\ee

The previous result imply that the time variation of $q$ is dominated by events 
which involve the rearrangements of regions of size which increase at least as 
$t^{\mu/3}$.  This type of behaviour (i.e.  a dynamical correlation length 
increasing as a power of time) has been seen in quenched disordered systems like 
spin glasses \cite{Rieger,Noi}.  It is quite evident that there is a change in 
the behaviour when the number particles involved in a typical rearrangement 
becomes of the same order of the sample size.  Independently from any 
theoretical speculations it is clear than when the variance $\Delta$ becomes 
comparable to $(1-q)^2$ the distribution of $q$ cannot be anymore Gaussian (the 
overlap cannot become greater than 1!) and we enter in a new regime.  Strong 
finite size effects are thus expected for sufficient large time, i.e.  for times 
which increase as $N^{1/\mu}$.  It is also quite likely that for sufficient 
large times a finite system reaches a sufficient low state such that further 
jumps are inhibited or happen on a much lager time scale \cite{Bo2}.

\begin{figure}[htbp]
  \epsfxsize=400pt\epsffile[22 206 549 549]{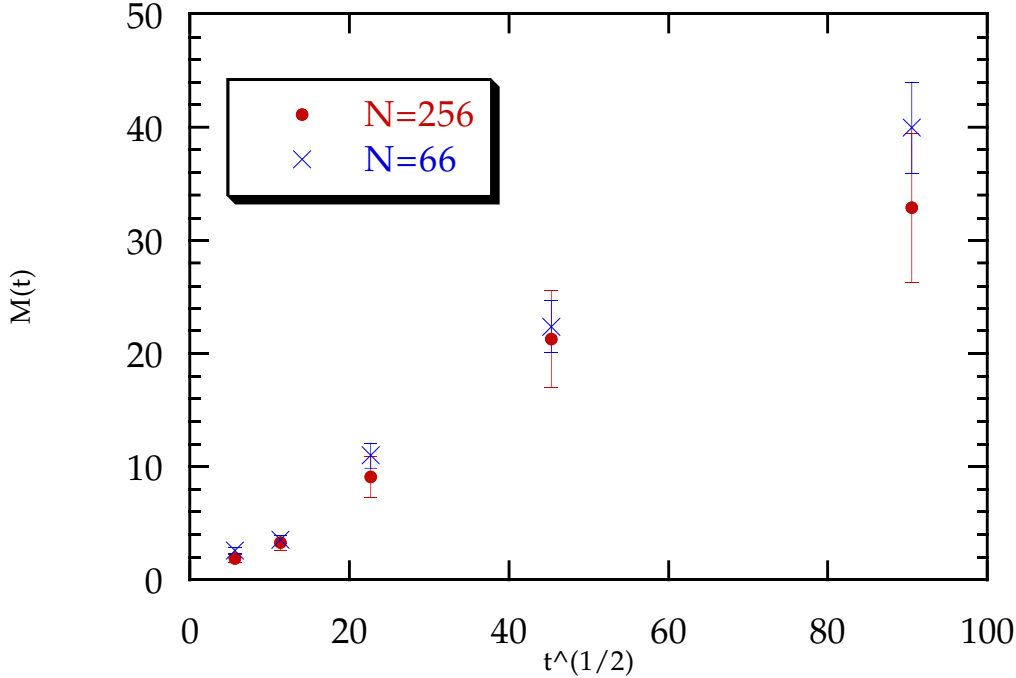}
\caption{The value of $M(t_w)$ as function of $t_w^{1/2}$ at $s=4$.}   
 \label{FIG5} \end{figure}

Summarizing the conclusion are this note are the following:
\begin{itemize}
\item Aging effects can be observed also at relatively short times in binary 
glasses.
\item Simple
aging is observed for a variation of about two order of magnitude in the waiting 
time with deviations with are at most 10\%.
\item It is suggested that there is a dynamical correlation length, that 
indicate the size of the regions which are collective rearranged, which diverges 
as a power of time.  (A direct study of the size of the rearranged regions in the 
equilibrium dynamics can be found in \cite{LAPA}).
\end{itemize}

Further numerical simulations are needed to decide if the small violations of simple
scaling are finite volume effects or they survive in larger samples. It would be
also interesting to study the temperature dependence of the effect and compare the
results with detailed theoretical predictions.

 \section*{Acknowledgments} I thank you G. Ciccotti and D. Lancaster for useful
discussions.

\end{document}